# Laser-ablation-assisted SF$_6$ decomposition for extensive and controlled fluorination of graphene

Jan Plšek*[a], Karolina Anna Drogowska, [a] Michaela Fridrichová, [a] Jana Vejpravová [b] and Martin Kalbáč**[a]

[a] Department of Low-dimensional Systems, J. Heyrovský Institute of Physical Chemistry, Academy of Sciences of the Czech Republic, v.v.i., Dolejškova 3, 18223 Prague 8 Czech Republic

[b] Department of Condensed Matter Physics, Faculty of Mathematics and Physics, Charles University, V Holesovickach 2, 180 00 Prague 8, Czech Republic

**Abstract**

We present a safe, clean, convenient, and easy-to-control method for extensive fluorination of graphene using laser-ablation-assisted decomposition of gaseous SF$_6$ molecules. The decomposition process is based on irradiation of a silicon target using a Nd:YAG laser (532 nm) in an SF$_6$ atmosphere. The reactive species produced in the plume above the silicon target are consequently responsible for fluorination of graphene placed in proximity to the plume. The applicability of the proposed method is demonstrated on fluorination of CVD-grown graphene on copper foil. Samples with different fluorination levels were evaluated using X-ray photoelectron spectroscopy and Raman spectroscopy. The influence of the applied number of laser pulses on the nature of the C–F bonds, fluorine, and sulphur concentrations, as well as graphene damage, is discussed. The method enables control of fluorine content by simple counting of laser pulses. We show that a fluorination level corresponding to a stoichiometry up to CF1.4 can be achieved. The fluorination does not lead to the formation of Cu–F in contrast to previously published approaches.

* Corresponding author. Tel.: +(420)266053545, E-mail: jan.plsek@jh-inst.cas.cz (Jan Plšek)
** Corresponding author. Tel.: +(420)266053804, E-mail:martin.kalbac@jh-inst.cas.cz (Martin Kalbáč)

## 1. Introduction

Fluorinated single-layered graphene (1-LG-F) represents a material with a wide range of applications. In particular, this material can be beneficial for electronic devices because its electronic properties can be tuned by the degree of fluorination from conductor to semiconductor to isolator [1–3]. Because 1-LG-F is only 1 nm thick and can be potentially ultra flat, one can consider applications to even include a tunnel barrier in spintronic devices [4] or for construction of BISFET etc. [5] Because the precursor of 1-LG-F is graphene, one can also consider patterned fluorination, which will lead to graphene/1-LG-F electronic devices. However, it is important to note that any advanced application will rely on high-quality samples, which are produced for example by mechanical exfoliation or chemical vapour deposition (CVD) synthesis. In addition, the fluorination method has to be vacuum compatible to avoid contamination. This requirement naturally disqualifies the so-called "fluorographene" or more precisely fluorographite materials made in large quantities by wet chemical methods.

So far, different fluorination methods (direct gas-fluorination, plasma fluorination, hydrothermal fluorination, and photochemical/electrochemical) [6] together with different chemical agents such as $F_2$ [7], $XeF_2$ [8–10], $CF_4$ [11], and $SF_6$ [12] have been employed in fluorination of single-layer graphene [6,13–15]. While the fluorination using $XeF_2$ is accelerated by the presence of defects [16], fluorination using $CF_4$ or $SF_6$ needs the assistance of radiofrequency plasma, which leads to the formation of F ions and radicals during the discharge. To achieve a stable condition, the RF plasma source has to work at optimized conditions (experimental geometry, applied power, exposure time, and pressure). This requirement limits the parameters of the plasma and consequently the tunability of the fluorination reaction. Moreover, the position of the sample with respect to the discharge as well as applied acceleration voltage can induce surface damage [17,18]. Nevertheless, using gentle sources of fluorine such as sulphur hexafluoride can provide a safer and convenient alternative to $F_2$ and $XeF_2$. Moreover, plasma etching using $SF_6$ is commonly employed in the semiconductor industry [19].

Here, we describe a plasma fluorination method based on a specific plasma generation process. Contrary to other direct plasma creation procedures usually based on microwave discharge, we employ molecular decomposition of $SF_6$ by irradiation of a silicon-based target [15,20,21]. Using a pulsed laser beam of a Nd:YAG laser focused on a silicon target, $SF_6$ gas was decomposed and graphene placed near the plume was fluorinated. This set-up provided a small and time-limited discharge region which enables precise control of the amount of fluorinating agents. We have examined in detail the degree of fluorination in different experimental conditions and have demonstrated that the proposed method can be used successfully for fluorination of graphene. A fluorination level corresponding to $CF_{1.4}$ stoichiometry can be achieved.

## 2. Experimental

The graphene was synthesized by CVD on a polycrystalline copper foil (~25 μm thick, 99.9% pure). Initially, the copper foil was annealed at 1,273 K under $Ar/H_2$ (50 sccm) for 20 min. Subsequently, methane was introduced for another 20 min, and finally, the samples were cooled from 1,273 to 300 K under an $Ar/H_2$ flow. It was shown previously that these preparation conditions lead almost exclusively to the formation of graphene singlelayers [22,23].

Graphene was fluorinated in a home-made system (Fig. 1). Prior to fluorination, graphene on copper foil was mounted on a simple sample holder in a small vacuum chamber (volume ~ 15 $cm^3$) made from a conflat CF 40 viewport and a CF 40 to KF 16 adapter flange, which can be closed using an angle valve. After sample mounting, the chamber was connected to a gas-handling system and pumped down using a rotary pump to a pressure of ~$3\times10^{-3}$ mbar. Afterward, the chamber was filled with $SF_6$ (Sigma-Aldrich, 99.75%) at a pressure of ~$7\times10^{-3}$ mbar and closed with a valve. Then, a Nd:YAG laser (Nano S 60-30, Litron Lasers, UK) beam was focused on the Si target. The laser spot (~300 $μm^2$) was located at a distance of ~3 mm from the sample plane, which was perpendicular to the target (Fig. 1). A certain number of laser pulses (532 nm, 20 mJ/pulse, 10 ns duration, 0.2–5 Hz repetition frequency) was applied and after that, the fluorination chamber was evacuated. Samples were transferred and stored in an ambient atmosphere.

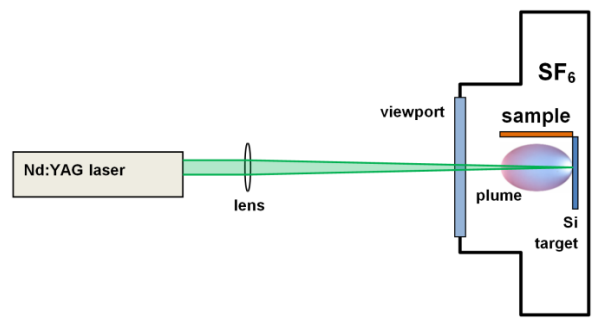

*Figure 1. Schematic diagram of graphene fluorination procedure by laser ablation of Si target in an $SF_6$ atmosphere.*

Immediately after fluorination, samples were characterized by XPS and Raman spectroscopy. The XPS measurements were performed in a VG ESCA3 MkII electron spectrometer with a base pressure better than $10^{-9}$ mbar. Al Kα radiation was used for the excitation of the electrons. The binding energies were referenced to the binding energy of Cu $2p_{3/2}$ (932.6 eV) electrons. The F/C ratios were calculated from the area under the C 1s photoelectron line and F 1s spectra after a Shirley background subtraction, assuming a homogeneous distribution of atoms, a Scofield photoionization cross-section, and a correction for the analyzer transmission function.

Raman maps were acquired using a WITec Alpha300 Raman spectrometer with a laser excitation wavelength of 633 nm and a laser power of 3 mW using a 50× objective. The acquisition time for each spectrum was 25 s and Raman maps comprised 40×40 spectra recorded in square 60×60 μm² areas containing several grains with different crystallographic orientations. Each Raman spectrum was fitted using Lorentzian line shapes in the WITec software for analysis of both D and G bands. Subtraction of a polynomial background over the entire spectral range was made prior to the peak fitting procedure.

SEM measurements were performed as a last experimental step using a Hitachi S4800 scanning electron microscope.

3. Results and discussion

To verify that the laser-ablation-assisted fluorination procedure is applicable to graphene, we examined fluorinated samples by XPS and Raman spectroscopy. At first, we probed possible sample contamination by XPS survey spectra (Fig. 2). Because we used a silicon target and the sample was placed close to the plume, silicon could be expected to be a highly probable surface contaminant. To undoubtedly exclude or confirm the presence of Si on the graphene, we used as-grown graphene samples on copper foil to prevent the introduction of any contaminant during the graphene transfer procedure. Figure 2 clearly shows two opposite cases: the survey spectra in Fig. 2A of a sample fluorinated using 300 pulses in an $SF_6$ atmosphere did not reveal the presence of any silicon; survey spectra depicted in Fig. 2B confirmed the significant amount of silicon on the sample although it was fluorinated also by 300 pulses. The difference in the preparation conditions was that the repetition rate of the Nd:YAG laser for the first sample was 5 Hz while for the second sample it was 0.5 Hz. Therefore, for further experiments, we used a repetition rate of 5 Hz only.

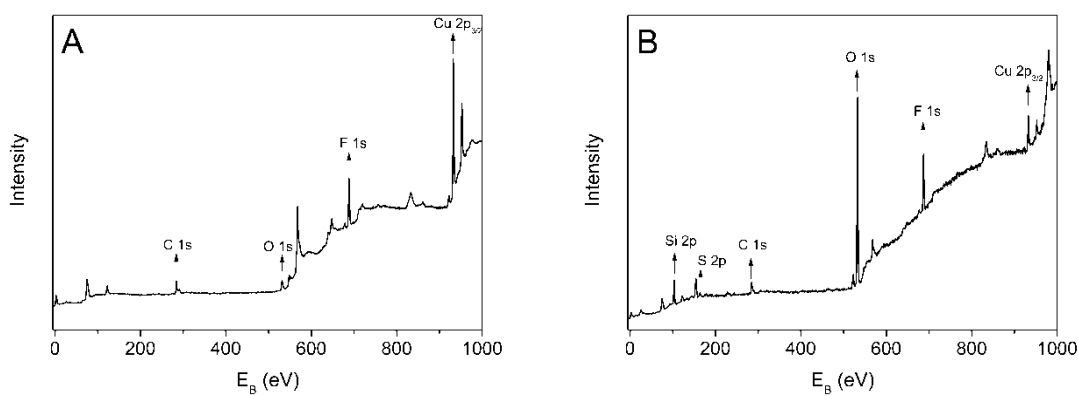

*Figure 2. Survey XPS spectra of sample fluorinated using 300 pulses in $7 \times 10^{-3}$ mbar of $SF_6$ at laser repetition rate of 5 Hz (A) and 0.5 Hz (B).*

Provided that graphene is not contaminated by silicon, the possibility of controlling the fluorination level would be crucial for successful application of the proposed method. The parameter that can be controlled in the easiest way is the number of laser pulses. Figure 3A illustrates that the fluorine atomic concentration increased up to approximately 60% after application of 150–240 laser pulses (Fig. 3A). A further increase of laser pulses caused a decrease of fluorine content. On the other hand, the sulphur amount (Fig. 3B) is the highest (S/C ~ 0.1) for the lowest number of laser pulses and it decreased with increasing number of laser pulses. After application of more than 240 laser pulses (Fig. 3B), sulphur was not detectable by XPS.

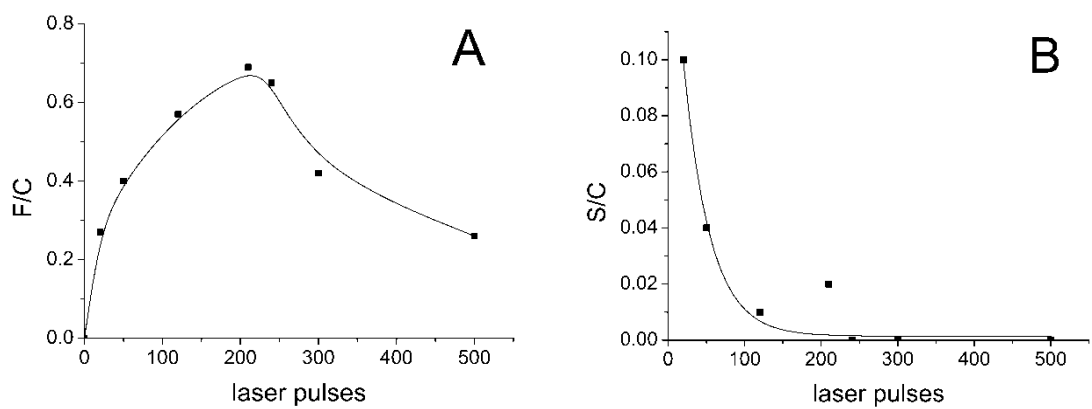

*Figure 3. Dependence of F/C (A) and S/C (B) atomic ratio on the number of applied laser pulses in 7×10–3 mbar of SF6 and laser repetition rate of 5 Hz. Black lines are eye guides only.*

The calculated stoichiometry for the highest fluorine coverage is approximately $CF_{1.4}$, which represents a several times higher degree of fluorination than the saturation fluorination level achieved for graphene on Cu foil fluorinated by $XeF_2$ [8,24,25]. The decrease of fluorine content for the higher number of laser pulses is presumably related to exhausting the $SF_6$ content in the chamber. A further increase of laser pulse number (1,000) even caused contamination of the sample with silicon. Significant coverage of the surface with silicon was also confirmed by SEM (Fig. S1C). It should be noted that a few silicon particles (droplets) can also be found on the sample fluorinated with 120 pulses (Fig. S1A) and slightly more for the sample fluorinated with 500 pulses (Fig. S1B). For pulsed laser deposition, it was suggested that contamination by droplets that originate from the target can be avoided using several methods,[26,27] including the "off-axis" method, which involves placing the substrate parallel to the plume. In our particular application of laser ablation, optimization of sample–target geometry can thus be further improved. Nevertheless, the overall amount of silicon was so small that it was not detectable by XPS. The high coverage by silicon after fluorination using 1,000 pulses can also explain the influence of the repetition rate on presence/absence of silicon on graphene. After probable depletion of $SF_6$ in the fluorination chamber, ablated Si species are not completely consumed by reaction with $SF_6$ and thus can be deposited on a nearby graphene sample. Depletion of $SF_6$ can be reached also when lower repetition rates are applied, which for the same number of laser pulses sufficiently increases the overall duration of laser ablation. Within this proposed mechanism, a decreasing F content (see Fig. 3A) after prolonged reaction can also be rationalized as a consequence of depletion of $SF_6$. Fluorine species that hit graphene at a lower pressure preserve more kinetic energy due to the small number of collisions with background gas. Furthermore, other energetic species that are present in the plume can also cause some desorption of adsorbed fluorine.

Generally speaking, fluorination of graphene by decomposition of $SF_6$ can be accompanied with co-deposition of sulphur [17,28]. We have detected sulphur for a lower number of laser pulses as well (Fig. 3B). The binding energy of the S 2p photoelectron line was ~164 eV (see Fig. S2), which corresponds to a C–S–C bond [29–33], which is weaker than a C–F bond. At higher fluorine coverage, the sulphur was however not found in the XPS spectra. We assume that sulphur is displaced by fluorine because the C–F bond is thermodynamically more stable.

Figure 4 shows examples of the F 1s XPS spectra for an increasing number of laser pulses during fluorination. Two components in the spectra of F 1s were observed. The major high-energy component at $E_b$ ~ 688.4 eV was assigned to the covalent C–F bond. The binding energy of the second, low-energy, component ($E_b$ ~ 686.5 eV) suggests less strong C–F bonding [34] which can be assigned to individual F atoms covalently bound to C in the basal plane [17]. With increasing fluorine coverage, the fraction of the low-energy component also increases (see Fig. 4). One can, therefore, conclude that during the early stage, fluorination proceeds preferentially at sites with lower coordination number. As the fluorination level increases, both components of F1s spectra were found to subsequently shift to a higher binding energy by approximately 1.5 eV (Figure S3). This shift indicates a decrease in the F–C bond length, which is directly related to an increase in covalency of the C–F bonding state [19],[34]. Interestingly, we did not observe the presence of F–Cu bonds such as for samples of graphene on Cu fluorinated by $XeF_2$ [24] or by RF plasma fluorination [17]. Because copper foil reacts very easily with fluorine, the absence of F-Cu bonds confirm that graphene samples is compact even after fluorination process.

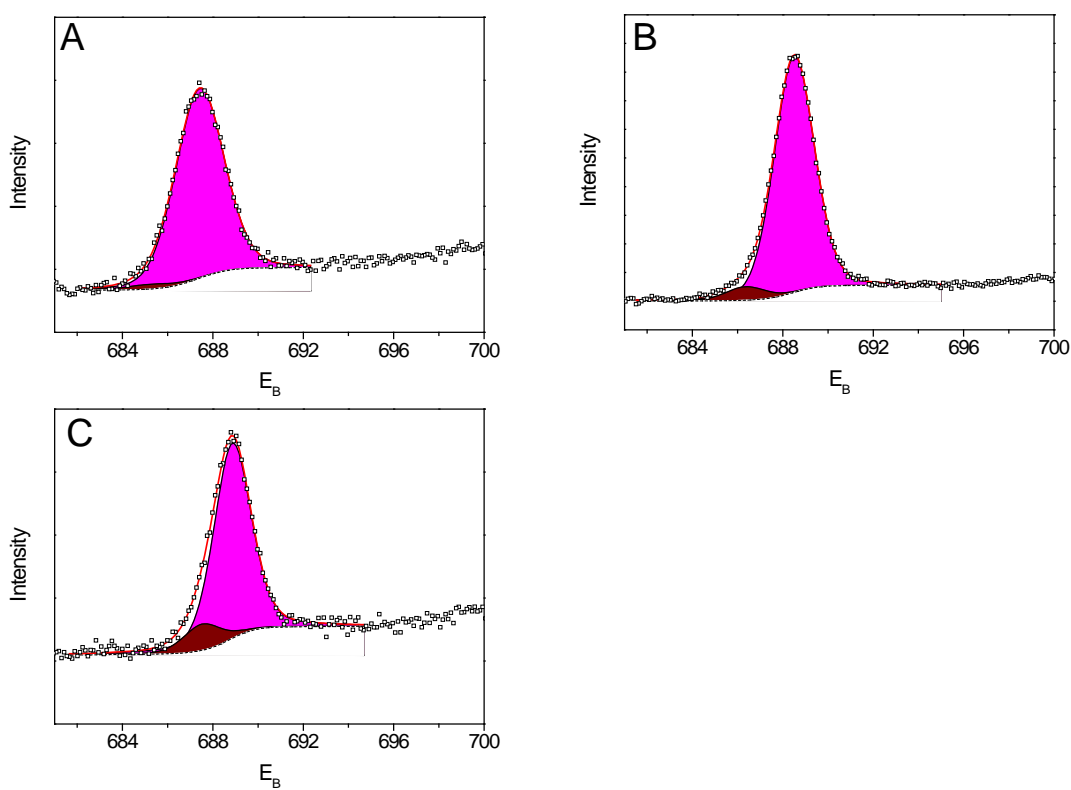

*Figure 4. Fitted spectra of F 1s photoelectrons obtained for samples fluorinated by 20 (A), 240 (B), and 500 (C) laser pulses in an atmosphere of $SF_6$ ($7\times10^{-3}$ mbar).*

The C 1s spectra of the fluorinated samples (Figure 5) were fitted by eight components. The two most intense components at 284.3 eV and 284.63 eV were fitted using simplified asymmetric peak line shapes and were assigned to coupled and decoupled $sp^2$ hybridized carbon states from the copper substrate [24],[35]. Similarly to highly oriented pyrolytic graphite [36], the simplified asymmetric peak line shape describes C 1s XPS line asymmetry and this asymmetry was suggested to be a consequence of the presence of several symmetric component peaks related to different defect states [37]. The two weak "shake-up" satellite peaks were fitted at a fixed separation from the sp$^2$ hybridized carbon peaks. Other components at ~286.5 eV, ~289 eV, ~291.2 eV, and ~293.5 eV were assigned as C–CF, C–F, CF$_2$, and CF$_3$. Assignment of CF species is in agreement with values reported previously [17],[38],[39]. However, similar as for F1s, we observed a shift of binding energy of individual components by ~1 eV (Figure S4). This shift is probably caused by increasing fluorine coverage in the nearest neighbors of all CF$_x$ species. Dependence of abundance of different CF components on the applied number of laser pulses shows that C–CF and CF components (Fig. 6A, 6B) are decreasing with increasing number of laser pulses, but CF$_2$ and CF$_3$ components (Fig. 6C and Fig. 6D) roughly exhibit similar behavior such as overall C/F ratio (see Fig. 3A).

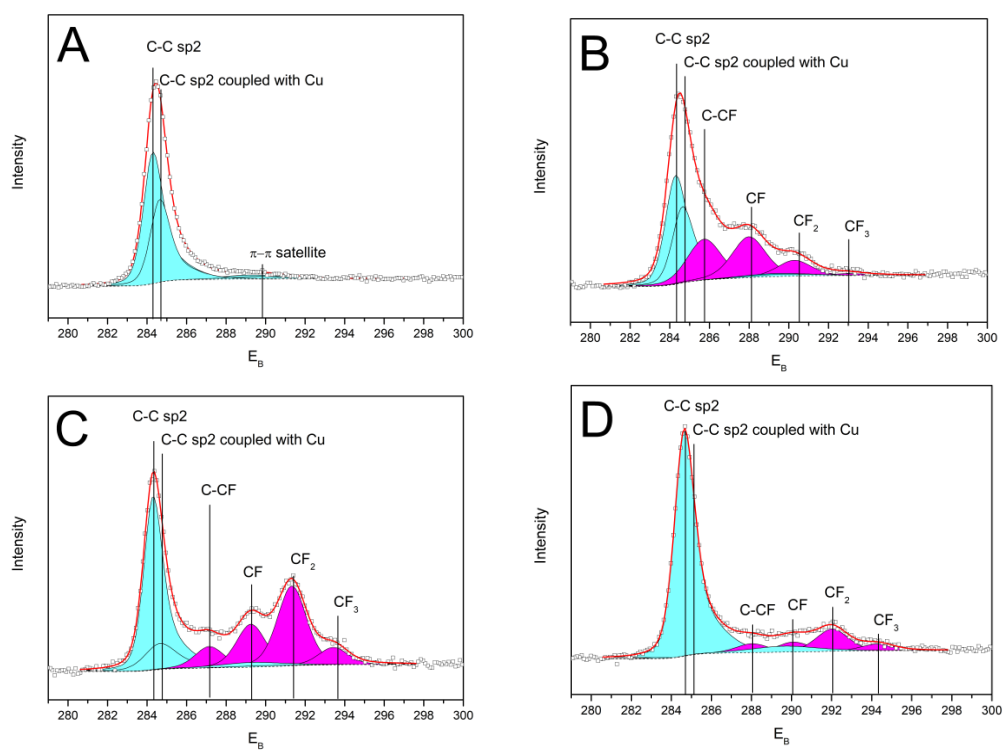

Figure 5. Fitted spectra of C 1s photoelectrons obtained for the non-fluorinated sample (A) and samples fluorinated by 20 (B), 240 (C), and 500 (D) laser pulses in $7\times10^{-3}$ mbar of $SF_6$.

Figure 7 illustrates the average Raman spectra calculated from Raman maps of graphene fluorinated by 50, 120, and 500 pulses in SF$_6$. All Raman spectra are similar, and their shapes are characterized by a considerably high intensity of the disorder-induced band (D peak) which increases with increasing number of laser pulses (Fig. 7B, Fig. S5). This increase is also connected with a substantial decrease of the 2D/G intensity ratio. For the highest number of laser pulses, a significant increase of the D peak width was also observed. Detailed mapping of selected areas resulted in an even distribution of the D/G ratio, which is in contrast to the samples fluorinated with $XeF_2$ [40]. Fluorination extent is thus not significantly influenced by the surface orientation of the underlying copper substrate.

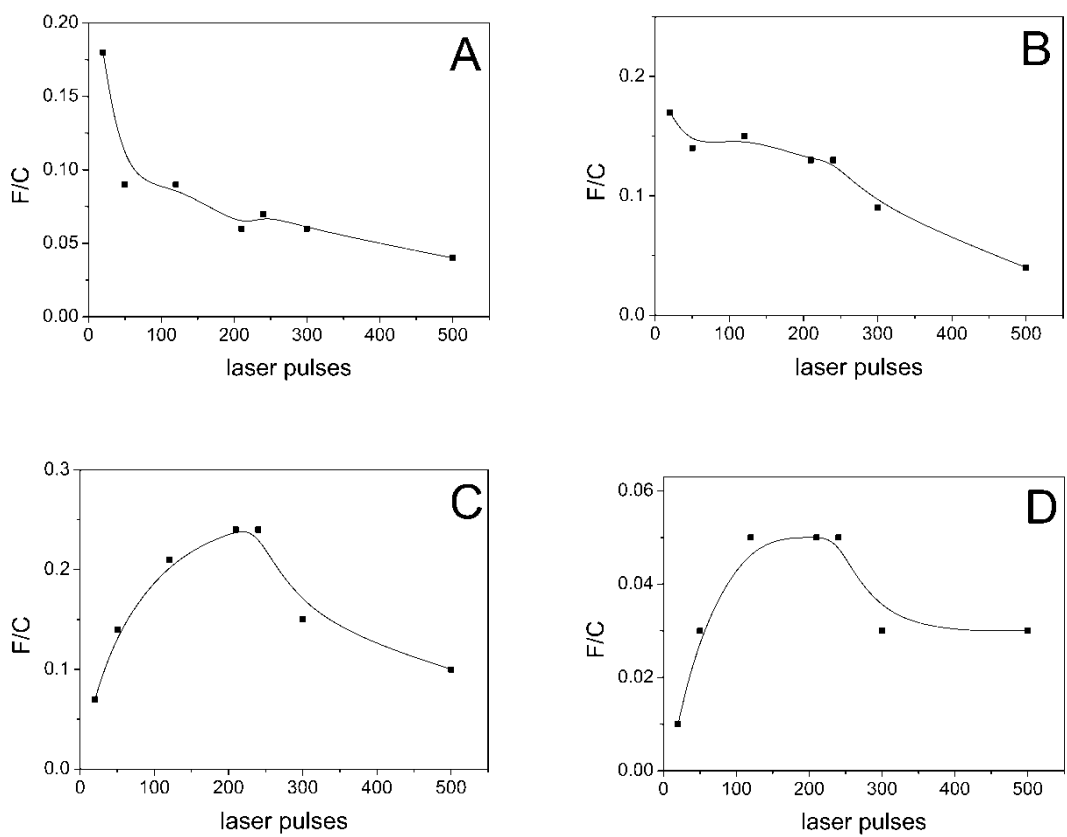

*Figure 6. The intensity ratio of C–CF (A), CF (B), CF$_2$ (C), and CF$_3$ (D) component of fitted C 1s XPS spectra as a function of applied laser pulses in $7\times10^{-3}$ mbar of SF$_6$ and laser repetition rate 5 Hz. Black lines are eye guides only.*

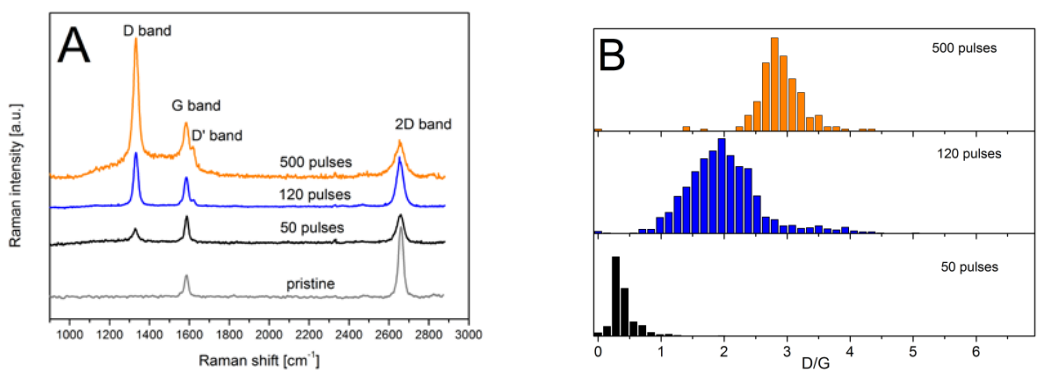

*Figure 7. (A) Average Raman spectra taken from entire Raman map area of graphene on copper foil fluorinated by 50, 120, and 500 pulses in $7\times10^{-3}$ mbar of $SF_6$ at a laser repetition rate of 5 Hz. (B) Histogram of the D/G ratio extracted from the entire measured area for graphene fluorinated by 50, 120, and 500 pulses in $7\times10^{-3}$ mbar of $SF_6$ and a laser repetition rate of 5 Hz.*

### 4. Conclusions

We present a new fluorination method based on laser-assisted plasma decomposition of gaseous $SF_6$ molecules. The reactive species are produced in the plume above the silicon target and are consequently responsible for fluorination of graphene placed in its proximity. The applicability of the proposed method was demonstrated using fluorination of CVD-grown graphene on copper foil. X-ray photoelectron spectroscopy and Raman spectroscopy confirmed the increasing fluorination extent with the number of applied laser pulses. The small volume (~15 cm$^3$) of the reaction chamber and low pressure of $SF_6$ was sufficient to reach the saturation fluorination value of graphene, which is approximately 60%, corresponding to the stoichiometry of $CF_{1.4}$. Several CF species were identified on graphene—CF, $CF_2$, and $CF_3$. Sulphur was identified only in the case of samples with low fluorine coverages. On the other hand, a higher number of laser pulses than the saturation value resulted in a decrease of fluorine content and, after depletion of $SF_6$, even deposition of silicon from the target was observed.

This new fluorination method preserves all advantages of current plasma fluorination methods. Using $SF_6$ as a gentle source of fluorine provides a safer, clean, convenient, and relatively easy-to-control method. The proposed new method furthermore minimizes the consumption of $SF_6$, which is a highly potent greenhouse gas. The small volume and simple construction of the reaction chamber are also very robust against damage by reactive species created during fluorination. The method offers a unique possibility to control the fluorination extent by the simple counting of laser pulses.

## Acknowledgement

The work was supported by the Czech Science Foundation project (18-20357S) and MEYS project LTC18039. We also acknowledge the assistance provided by the Research Infrastructure NanoEnviCz, supported by the Ministry of Education, Youth and Sports of the Czech Republic under project no. LM2015073 and project no. CZ.02.1.01/0.0/0.0/16_013/0001821.

**Graphical abstract**

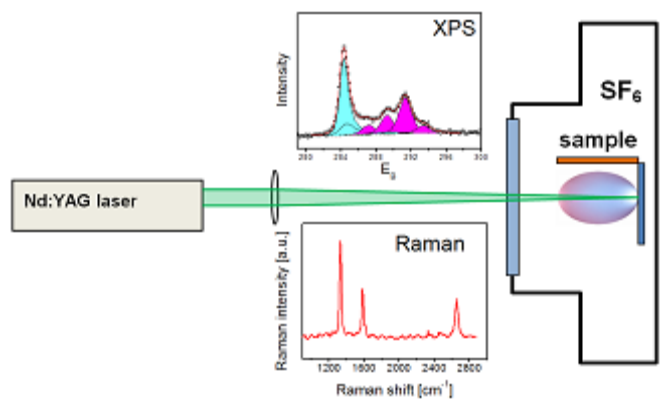